\documentclass[11pt]{article}
\usepackage{blois,epsfig}

\def\la{\mathrel{\mathpalette\fun <}}

\def\fun#1#2{\lower3.6pt\vbox{\baselineskip0pt\lineskip.9pt
\ialign{$\mathsurround=0pt#1\hfil##\hfil$\crcr#2\crcr\sim\crcr}}}

\def\Journal#1#2#3#4{{#1} {\bf #2}, #3 (#4)}

\def\PLB{{\em Phys. Lett.}  B}
\def\PRL{\em Phys. Rev. Lett.}
\def\PRD{{\em Phys. Rev.} D}

\def\be{\begin{equation}}
\def\ee{\end{equation}}
\def\bea{\begin{eqnarray}}
\def\eea{\end{eqnarray}}

\begin{document}
\vspace*{4cm}
\title{FOURTH QUARK-LEPTON GENERATION AND PRECISION MEASUREMENTS}

\author{ M.I. VYSOTSKY }

\address{ITEP, Moscow, Russia}

\maketitle \abstracts{ Precise measurements of $Z$-boson parameters and
$W$-boson and $t$-quark masses put strong constraints on non
$SU(2)\times U(1)$ singlet New Physics. We demonstrate that one extra
generation passes electroweak constraints even when all new particle
masses are well above their direct mass bounds. My talk is based on a
recent paper \cite{1}.}

\section{Decoupling and nondecoupling}

Let us start from Quantum Electrodynamics and consider heavy $t$-quark
contributions to physical observables. Contribution of a diagram with
top quark propagating in a loop to fine structure constant $\alpha$ is
divergent and not extractable from data. $t$-quark contribution to muon
anomalous magnetic moment is suppressed as $\alpha^2(m_\mu/m_t)^2$, so
it decouples.

What about electroweak theory? $t$-quark contribute into $K-\bar K$,
$B-\bar B$ transition amplitudes $\sim f^4/m_t^2 = G_F^2 m_t^2$, where
$f$ is Yukawa coupling of top with higgs boson, $f\sim m_t/\eta$, so it
does not decouple. Even if new generations mixing with light
generations is small (vanishing contributions to $K-\bar K$, $B-\bar B$
mixing) they contribute to $Z$ and $W$ polarization operators.
Resulting contributions to physical observables are finite and do not
decouple. This is precisely the reason why considerable part of phase
space (masses of new quarks and leptons) is excluded by precision data.

\section{SM fit}

The precision data fit within Standard Model with three generations
performed by LEPTOP code is presented in \cite{1}. The data set
contains $Z$-boson decay parameters, $W$-boson and $t$-quark masses and
the value of running fine structure constant at $q^2 = m_Z^2$.

The quality of the fit is characterized by $\chi^2/n_{d.o.f.} = 18/12$
and is reasonably good. Higgs boson is predicted to be light, $m_H =
84^{+32}_{-24}$ GeV; at $\mu = m_Z$ the value of QCD coupling constant
$\hat\alpha_s = 0.184(27)$.

\section{Fits with extra generation}

Introducing the fourth generation we assume that the mixing with three
known generations is small. We fix the value of the sum of new quark
masses $m_U + m_D = 600$ GeV to avoid Tevatron direct search bounds and
the value of the charged lepton mass $m_E = 200$ GeV (well above LEP II
bound) and look for a minimum of $\chi^2$ for the fixed values of $m_H$
varying neutral lepton mass and the difference of Up- and Down-quark
masses. From the width of $Z$ we know that neutral lepton should be
heavy, $m_N > 50$ GeV. The results of fits for $m_H = 120$ GeV and $m_H
= 1000$ GeV are shown on Figures 1 and 2.

\begin{figure}[!htb]
\centering \epsfig{file=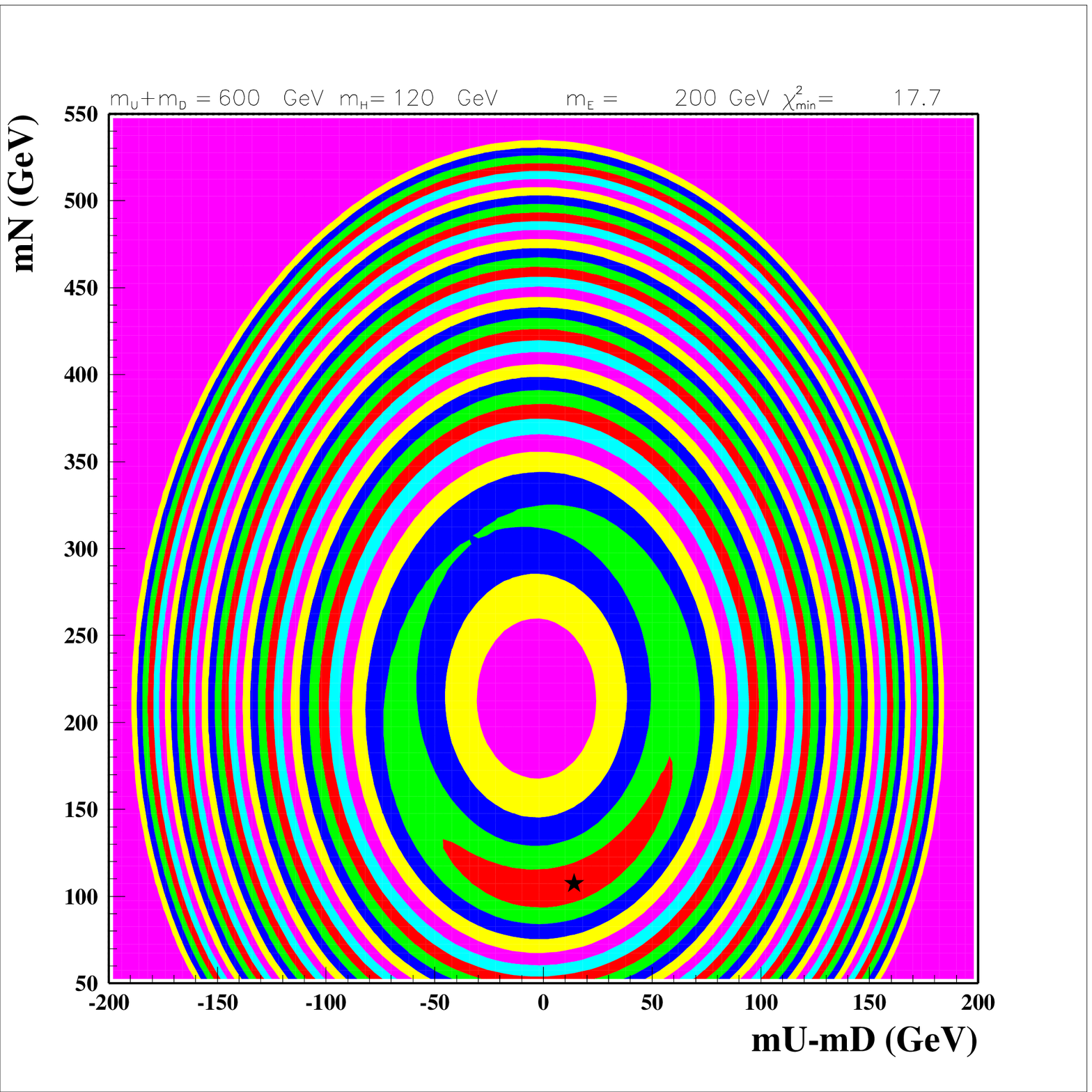,width=5cm} \caption{\em Exclusion plot
on the plane $m_N$, $m_U - m_D$ for fixed values $m_H = 120$ GeV, $m_E
= 200$ GeV, $m_U + m_D = 600$ GeV. $\chi^2$ minimum shown by the star
corresponds to $\chi^2/d.o.f. = 17.7/11$.The borders of the regions
show domains allowed at the level $\Delta\chi^2 = 1,4,9,16$ etc. }
\label{WW2Fermi}
\end{figure}

\begin{figure}[!htb]
\centering \epsfig{file=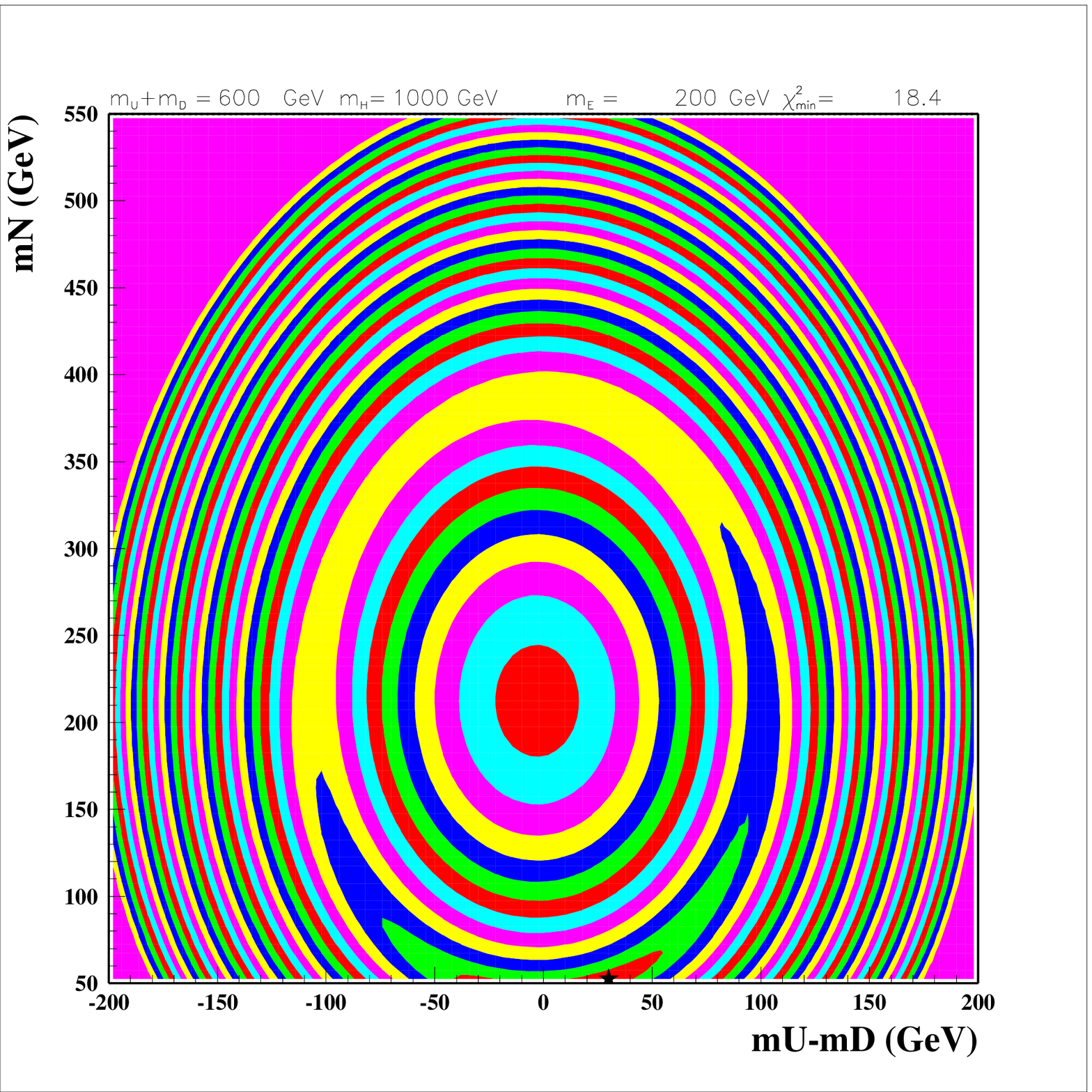,width=5cm} \caption{\em The same as
Fig. 1 for $m_H = 1000$ GeV. $\chi^2$ minimum shown by the star
corresponds to $\chi^2/d.o.f. = 18.4/11$. } \label{WWFermi}
\end{figure}

We see that the quality of fits is almost the same as in the case of
three generations.

\section{$\boldmath S$, $T$, $U$ versus $\boldmath S_i$, $T_i$, $U_i$}

These variables are frequently used to determine if a given new physics
model passed constraints of precision measurements.

\newpage

\begin{center}

{\bf Table}

\bigskip

\begin{tabular}{|c|c|c|c|c|c|c|c|}
\hline $m_H$ & & $T_i$ & $T$ & $S_i$ & $S$ & $U_i$ & U \\
\hline 120 & $m_U = 310$, $m_D = 290$ & 0.02 & 0.02 & 0.15 & 0.15 & 0 &
0 \\ \cline{2-8} & $m_N = 120$ , $m_E = 200$ & 0.11 & 0.11 & -0.01 &
-0.01 & 0.02 & 0.01 \\ \hline 1000 & $m_U = 315$, $m_D = 285$ & 0.05 &
0.05 & 0.15 & 0.15 & 0 & 0 \\ \cline{2-8} & $m_N = 53$, $m_E = 200$ &
0.27 & 0.36 & -0.19 & -0.13 & 0.16 & 0.10 \\ \hline
\end{tabular}

\bigskip

{\em Quark and lepton contributions to $S, T, U$ and $S_i, T_i, U_i$ at
the points of $\chi^2$ minimum in Figures 1 and 2. All masses are in
GeV.}

\end{center}

\bigskip

In the Table values $S$, $T$ and $U$ for the points of $\chi^2$ minimum
in Figures 1 and 2 are presented. Definitions of $S$, $T$ and $U$ from
the review paper \cite{2} are used. Also the values $S_i$, $T_i$ and
$U_i$ where ``$i$'' means ``improved'' are presented. While used in
\cite{2} definitions of $S$, $T$ and $U$ correctly describe
contributions of heavy new particles, $(m_{n.p.}/M_Z)^2 \gg 1$,
quantities $S_i$, $T_i$ and $U_i$ can be used for $m_{n.p.} \la M_Z$ as
well (this improved set was introduced in \cite{1} where it was
designated as $S'$, $T'$ and $U'$). For the case of light higgs $m_H =
120$ GeV $S = S_i = 0.14$, $T = T_i = 0.13$ and both $U_i$ and $U$ are
almost zero. In Fig. 10.4 of \cite{2} one can see that both Standard
Model point $S = T = 0$ and just described new physics point are on the
border of the allowed $1\sigma$ domain for $m_H = 117$ GeV. This
confirms the statement about equally good data description by SM and by
the fourth family extension. For the case of heavy higgs $m_H = 1000$
GeV lepton contributions to $S$, $T$ and $U$ differs from that to
$S_i$, $T_i$ and $U_i$. Summing quark and lepton contributions we get:
$S = 0.02$, $T = 0.41$, $U = 0.10$ and the $(S,T)$ point is outside
$1\sigma$ allowed domain for $m_H = 1000$ GeV from Fig. 10.4. However
taken into account improved values $S_i = -0.04$, $T_i = 0.32$, $U_i =
0.16$ we see, that $(S_i, T_i)$ point is at the border of $1\sigma$
allowed domain for $m_H = 1000$ GeV from Fig. 10.4 as it should be
according to the analysis of Section 3. Similar statements can be found in \cite{43}.

\section{Higgs properties with the fourth generation}

Loops with $U$- and $D$-quarks contribute to $gg\to H$ production
amplitude. This helps CDF to exclude larger interval of $H$ masses than
in the case of 3 generations \cite{3}. $H$ decay branching ratious are
modified as well \cite{4}.

Additional heavy fermions generate negative loop correction to the
Higgs boson potential which can make it unbounded from below. The
allowed domains of fermion and Higgs boson masses from \cite{5} are
shown on Fig.~3. They depend on the ultraviolet cut-off $\Lambda$ at
which new physics comes into play. One can see that in case the fourth
generation exists we should have another type of relatively light
($\Lambda < 10^5$ TeV) new physics as well.

%\begin{center}

%\bigskip

%\includegraphics[width=.6\textwidth]{MHM4.pdf}

%\end{center}

\begin{figure}[!htb]
\centering \epsfig{file=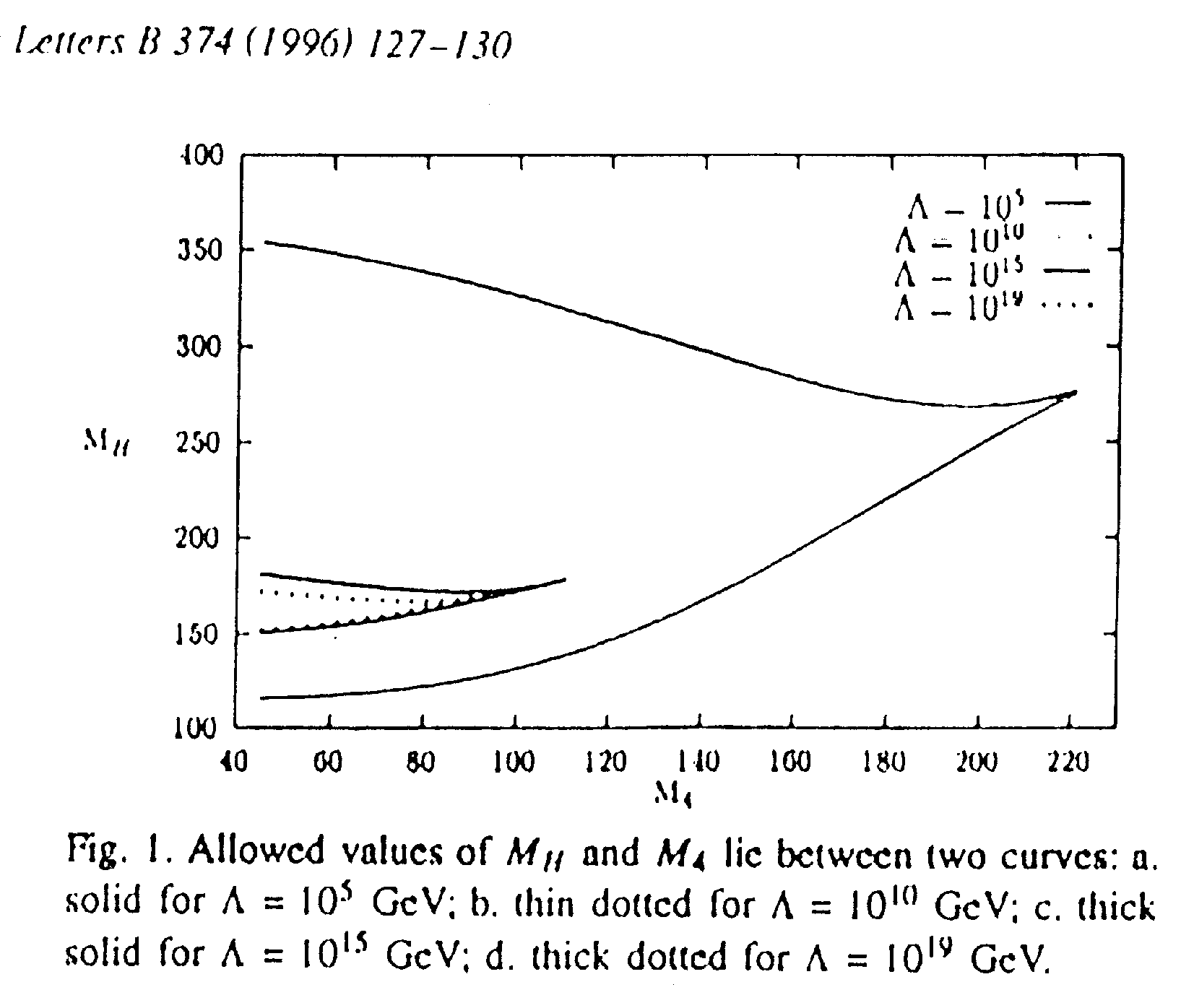,width=10cm}
\caption{\em Bounds from
the stability of Higgs potential and absence of the Landau pole. }
\label{WFermi}
\end{figure}

\section{Conclusion}

One extra quark-lepton generation is not excluded by electroweak
precision data.

The quality of fit for one extra generation is the same as that for
Standard Model for certain values of new particle masses.

In the case of the fourth generation the upper bound on higgs mass from
Standard Model fit is removed. Higgs production cross-section in gluon
fusion is enhanced while decay branching ratios are modified in the
case of extra generations.

The fourth generation can be used to explain possible deviations from
the Standard Model predictions for CP violation in $B$-meson decays
\cite{6}, \cite{7}, \cite{8}. However presently I cannot find experimentally
proved deviation which can not be attributed to the effects of strong
interactions.

I am grateful to my coauthors Victor Novikov and Alexander Rozanov.

I am grateful to the organizers and in particular to Jean Tran Than Van,
Blazenka Melic and Aleksander Khodjamirian
for the possibility to participate in a very interesting and exciting
conferences.

\end{document}